# Chiral superfluorescence from perovskite superlattices at room temperature

Qi Wei[1#], Jonah S. Peter[2,3,4#], Hui Ren[1#], Weizhen Wang[1], Luwei Zhou[1], Qi Liu[1], Stefan Ostermann[2], Jun Yin[1], Songhua Cai[1], Susanne F. Yelin[2]✉, Mingjie Li[1,5,6]✉

1. Department of Applied Physics, The Hong Kong Polytechnic University, Hung Hom, Kowloon, Hong Kong, China
2. Physics Department, Harvard University, Cambridge, Massachusetts 02138, USA
3. Biophysics Program, Harvard University, Boston, Massachusetts 02115, USA
4. Department of Chemistry and Chemical Biology, Harvard University, Cambridge, Massachusetts 02138, USA
5. Shenzhen Research Institute, The Hong Kong Polytechnic University, Shenzhen, Guangdong, 518057, China
6. Photonics Research Institute, The Hong Kong Polytechnic University, Hung Hom, Kowloon, Hong Kong, China

Correspondence: syelin@g.harvard.edu; ming-jie.li@polyu.edu.hk

[#]These authors contribute equally to this work.

**Abstract**

Superfluorescence (SF) is the collective emission of intense, coherent light from an interacting ensemble of quantum emitters[1-4]. While SF has been observed in several solid-state materials[5-8], the spontaneous generation of circularly polarized SF from chiral materials (chiral SF) has not been realized[9,10]. Here, we report the first observation of chiral SF originating from edge states in large-area (>100 μm × 100 μm) vertically aligned chiral perovskite superlattices at room-temperature. Theoretical quantum optics calculations describe the transition from initially unpolarized, incoherent spontaneous emission to a coherent chiral SF state, quantitatively reproducing both the experimentally observed generation of circular polarization (up to ~14%) and its reversal in sign with opposite material handedness. Moreover, we show that both the intensity and the degree of circular polarization of chiral SF can be modulated by a weak magnetic field, enabling precise control over solid-state quantum light emission at room temperature. Our findings demonstrate an interplay between chirality and many-body quantum coherence, thereby revealing promising new directions for chirality-controlled quantum-optical applications.

## Main text

Superfluorescence (SF) and superradiance describe remarkable quantum optical phenomena in which the light radiated from an ensemble of quantum emitters is enhanced through cooperative light-matter interactions[1-4]. Although the terms are often used interchangeably, superradiance traditionally refers to radiation from a correlated initial state[11], whereas SF originates from an initially uncorrelated state that develops a spontaneously enhanced dipole moment through interactions with the electromagnetic vacuum (Supplementary Note 1). As such, SF is a fundamentally quantum mechanical effect that offers unique insight into many-body correlations and entanglement dynamics in photonic systems[12-15]. The development of strongly superfluorescent quantum materials could thus drive significant advancements in optoelectronics and quantum technologies, including ultrafast quantum memories, high-speed optical interconnects, and scalable quantum information processing architectures[16-18].

In recent years, chiral materials have emerged as promising platforms for manipulating correlated quantum dynamics. The discovery of the chirality-induced spin selectivity (CISS) effect has stimulated the development of next-generation spintronics devices, with relevance to both classical and quantum information processing[19-23]. More recently, a photonic analog of CISS arising from *chiral* SF—where the circular polarization of the superflourescent light is determined by the handedness of the chiral material—was predicted theoretically[9,10], synergizing the transformative aspects of coherent quantum optics with those of chiral materials. While SF has been observed in several materials, including cryogenically cooled InGaAs quantum wells under strong magnetic fields[5], perovskite quantum dots at low temperatures[6], and quasi-2D hybrid perovskite thin films at high temperatures[8,24], its realization in chiral solid-state architectures remains elusive. Moreover, despite its revolutionary potential for both photonic and quantum optical applications, chiral SF (or superradiance) has not been demonstrated in any experimental system. The development of scalable fabrication techniques for uniform and reproducible solid-state SF also remains a significant hurdle for widespread implementation.

In this study, we report the observation of room temperature chiral SF originating from the edge states of vertically aligned quasi-2D hybrid organic–inorganic perovskite superlattices. We demonstrate that the chirality of the superlattices, achieved through chirality transfer via chiral ligands[25], enables strong chiral SF with a degree of circular polarization (DCP) of up to ∼14%.

Our quantum optics calculations quantitatively reproduce the observed DCP amplification at high excitation density, as well as its reversal in sign with opposite material handedness. Moreover, we show that a weak external magnetic field (< 0.5 T) can further enhance both the chiral SF intensity and the DCP. Our results demonstrate an intimate connection between chirality and macroscopic quantum coherence in perovskite superlattices and open promising avenues for quantum spin-optical applications[26].

**Chiral perovskite superlattices and time-integrated PL**

We developed three types of quasi-2D ($n > 1$) perovskite superlattices to probe solid-state SF. Each material follows the formula $L_2MA_{n-1}Pb_nI_{3n+1}$ (Fig. 1a), where $n$ is the number of inorganic octahedral layers per quantum well, and the quantum well spacer L is either the achiral ligand phenethylamine (PEA), PEA mixed with the left-handed ligand S-(+)-$\alpha$-methyl benzylamine (SMBA), or PEA mixed with the right-handed ligand R-(+)-$\alpha$-methyl benzylamine (RMBA). Large-area (> 100 µm × 100 µm) "crisscross network"-structured quantum well superlattices were grown vertically on a $MAPbBr_3$ single-crystal substrate (Fig. 1b, Extended Data Figs. 1-2, Supplementary Note 2, and Supplementary Figs. 1-2). Superlattice elements within each perovskite sample measure ~200–300 nm in width, ~0.5–5 µm in length, and ~1.5 µm in height (Fig. 1c). Selective precursor–substrate bonding ensures vertical alignment of the quantum wells[27]. Top-down and cross-sectional scanning transmission electron microscopy (STEM) images clearly resolve the quasi-2D quantum-well superlattices with a dominant $n = 3$ phase and a minor $n = 2$ component (Fig. 1d). Both STEM and synchrotron-based grazing-incidence wide-angle X-ray scattering (Extended Data Fig. 3) further confirm the long-range periodic order of the vertically aligned superlattices.

Our superlattices are excited using a 550 nm linearly polarized ~50 fs laser pulse tuned above the material's bandgap. Due to their isotropic crisscross morphology, the superlattices are insensitive to the polarization state of the pump laser and exhibit no preferential absorption or propagation of circularly polarized light (Supplementary Note 3, Extended Data Fig. 4). At high pump fluence, the excitation density increases, and the dipoles interact coherently through shared electromagnetic modes—eventually leading to the formation of a giant coherent dipole. In our chiral superlattices, this process gives rise to a collectively enhanced circularly polarized dipole resulting in chiral SF (Fig. 1e).

Pump fluence-dependent photoluminescence (PL) spectra were obtained by integrating the time-dependent fluorescence intensity under repeated femtosecond laser excitation with a 1 kHz repetition rate. At low pump fluence, the superlattices exhibit a broad spontaneous emission peak at ~690 nm originating from the edge states of the vertically oriented quasi-2D perovskites[28,29] (Extended Data Fig. 5, Supplementary Note 4, Supplementary Figs. 3–6). The superlattices also show signatures of coherent coupling, with an enhanced radiative rate under resonant excitation (Supplementary Note 5, Supplementary Figs. 7–10). Above a threshold pump fluence, the PL spectra become sharp and narrow-band, and the peak intensity increases superlinearly: a key signature of SF (Fig. 1f). As estimated from time-integrated PL spectra, the SF thresholds are $70 \pm 10$ μJ $cm^{-2}$ for the left-handed SMBA sample (peak intensity at 705 nm), $95 \pm 10$ μJ $cm^{-2}$ for the right-handed RMBA sample (710 nm), and $140 \pm 10$ μJ $cm^{-2}$ for the achiral PEA sample (690 nm) (Supplementary Fig. 11). The PL spectra demonstrate relatively high photostability both below and above SF threshold (Supplementary Fig. 12).

**Superfluorescent characteristics and dynamics**

Coherent SF exhibits distinctive properties that differentiate it from incoherent radiative processes with similar time-integrated spectra. For instance, the SF observed in our samples is directional with peak intensity normal to the substrate. This directionality persists for both stripe and point beam excitation, in contrast to the amplified spontaneous emission characteristics typically observed in conventional perovskite systems[30,31] (Supplementary Note 6, Supplementary Figs. 13–14). The inset of Fig. 1f shows a spatial interferogram of the SMBA perovskite superlattices pumped above the SF threshold, as measured using a Michelson interferometer. Above threshold, the interference fringes are clearly visible across distances up to 1 mm, indicating the buildup of long-range spatial coherences (Supplementary Note 7). Below threshold, no interference fringes are observed (Supplementary Fig. 15). Moreover, time-resolved SF dynamics exhibit characteristic oscillations known as Burnham-Chiao ringing (Fig. 2a,b), stemming from the coherent re-absorption and re-emission of photons within the excited medium after the initial SF burst. This provides a key signature of the coherent coupling between emitters in the SF state[32]. Similar ringing behavior is observed in the PEA and RMBA superlattices (Extended Data Figs. 6-7), and similar SF thresholds are inferred from time-resolved PL (TRPL) as for time-integrated PL.

Importantly, the measured peak SF intensity $I_{\mathrm{SF0}}$ also follows the theoretically predicted quadratic scaling, $I_{\mathrm{SF0}} \propto N^2$, where $N$ is the initial number of photoexcited dipoles (Supplementary Note 1). This scaling is unique to SF and contrasts with the linear scaling associated with incoherent spontaneous emission ($I_{\mathrm{SE}} \propto N$). To verify this relationship, we first determined the scaling between pump fluence $P$ and dipole density $N$ in the spontaneous emission regime using TRPL measured below the SF threshold (Fig. 2b). For the SMBA superlattices, the spontaneous emission scales as $I_{\mathrm{SE0}} \propto P^{1.55 \pm 0.06}$, while the SF intensity scales as $I_{\mathrm{SF0}} \propto P^{3.16 \pm 0.07}$. Together, these measurements lead directly to the quadratic relationship $I_{\mathrm{SF0}} \propto N^{2.04 \pm 0.09}$ (Fig. 2c). Similar scaling behavior is observed for the PEA and RMBA superlattices (Extended Data Figs. 6-7).

Another characteristic feature of SF is the delay time $\tau_D$ between the onset of initially incoherent spontaneous emission and the coherent SF peak. This delay decreases with increasing pump fluence as more photoexcited dipoles are recruited for collective enhancement. Transient absorption (TA) spectroscopy probed at the SF wavelength performed on the superlattices transferred to a transparent quartz substrate reveals this important feature of SF (Fig. 2d, Supplementary Fig. 16). Below the SF threshold, we observe a rapid population buildup (~ 1 ps) followed by slow population decay (essentially constant throughout the measurement window). Above the SF threshold, a characteristic delay period ($\tau_D$) precedes the rapid population decay from the SF burst. The $\tau_D$ values extracted from both TA and TRPL measurements follow the theoretically predicted scaling for SF, $\tau_D = (1/\Gamma)\ln(N)/(N+1)$ (Ref 33) (Fig. 2b, right and Supplementary Fig. 17), where $\Gamma = 1/1500$ ps$^{-1}$ is the radiative rate extracted from spontaneous emission dynamics (Supplementary Fig. 6c).

In our superlattices, room temperature SF is enabled by the relatively long coherence time $T_2$ of the fluorescent edge states, which exceeds the SF delay time and pulse duration over the full range of pump fluences. Transient four-wave-mixing measurements show that the edge state coherence time decreases from ~ 72 ps at low pump fluence to ~ 11 ps at high fluence for the SMBA superlattices (Supplementary Fig. 18). The observation of long coherence times exceeding 10 ps is consistent with previous reports of high-temperature SF in similar quasi-2D hybrid perovskites and may arise from suppressed exciton-phonon coupling via polaron dressing[8,24]. In our samples, additional protection against decoherence is likely related to exciton localization in deep edge states[28], which are known to suppresses inelastic relaxation pathways[34] (Supplementary Note 8).

**Circularly polarized chiral SF**

Having confirmed the superfluorescent nature of the narrow band emission at ~700 nm through spectral and dynamical assessments, we next investigated the effect of chirality on the polarization of the SF burst (Fig. 3a). Circular dichroism (CD) spectra confirm the successful transfer of chirality from the (S/R)MBA ligands to the perovskite superlattices (Supplementary Note 9, Supplementary Figs. 19-20). Notably, the use of a linearly polarized pump laser avoids the direct injection of polarized states through circularly polarized excitation, which would otherwise obscure the spontaneous generation of circular polarization via cooperative SF. Remarkably, we observe strong circularly polarized SF from the chiral perovskite superlattices at room temperature (Fig. 3b). Consistent with theoretical predictions, the intensity difference between the left-circularly polarized (LCP) and right-circularly polarized (RCP) emissions changes sign between the chiral RMBA and SMBA configurations. The degree of circular polarization at peak SF intensity is defined as $\mathrm{DCP}_0 = \frac{I_{\mathrm{LCP0}} - I_{\mathrm{RCP0}}}{I_{\mathrm{LCP0}} + I_{\mathrm{RCP0}}} \times 100\%$, where $I_{\mathrm{LCP0}}$ and $I_{\mathrm{RCP0}}$ are the peak LCP and RCP intensities extracted from TRPL spectra (Fig. 3c, Supplementary Fig. 21). In the SF regime, we report a $\mathrm{DCP}_0$ of up to 13.5% for the SMBA and 13.8% for the RMBA superlattices (data points in Fig. 3d). Comparable values are obtained for the time-integrated PL intensities at peak wavelength (Supplementary Fig. 22). By contrast, the DCP of spontaneous emission measured below the SF threshold is negligible at room temperature (~0.2%, Supplementary Fig. 23), consistent with previous work on chiral 2D/quasi-2D perovskites synthesized with same chiral ligands[35-37]. Additionally, no circularly polarized SF is observed in the achiral PEA superlattices, and no SF is observed in a conventional quasi-2D chiral perovskite thin film (Supplementary Figs. 24-25).

The chiral SF observed here is a many-body generalization of the single-photon chiral SF predicted theoretically in chiral dipole arrays[9,10]. The spontaneous generation of circular polarization in the SF regime can be explained theoretically by extending the usual single-mode SF laser model[33] to include geometry and polarization effects. Here, we consider a simplified one-dimensional lattice model that captures the screw axis symmetry of our vertically-aligned chiral superlattices. The dipole operator for a two-level emitter at position $z_j$ along the helical axis is $\mathbf{p}_j = \wp_j(\sigma_j^{ge} + \sigma_j^{eg})$, where $\sigma_j^{eg}$ and $\sigma_j^{ge}$ are raising and lowering operators, respectively, for the ground and excited states. The transition dipole moment twists along the screw axis and can be written as

$$\wp_j = \wp\left(\cos(pz_j)\hat{\mathbf{x}} + \sin(pz_j)\hat{\mathbf{y}}\right) = \frac{\wp}{\sqrt{2}}\left(e^{ipz_j}\hat{\mathbf{e}}_- + e^{-ipz_j}\hat{\mathbf{e}}_+\right), \quad (1)$$

where $\wp$ is the magnitude of the transition dipole moment, $2\pi/p$ is the lattice spacing (Fig. 3a) and $\hat{\mathbf{e}}_\pm = (\hat{\mathbf{x}} \pm i\,\hat{\mathbf{y}})/\sqrt{2}$ are the unit vectors of circular polarization. Accounting for both polarizations, the single mode cavity field along this axis is given by

$$\mathbf{E}(z) = \sum_{\lambda=\pm} \mathcal{E}\left(a_\lambda e^{ik_0 z}\hat{\mathbf{e}}_\lambda + a_\lambda^\dagger e^{-ik_0 z}\hat{\mathbf{e}}_\lambda^*\right). \quad (2)$$

Here, $\mathcal{E}$ is the field amplitude, $a_\lambda^\dagger$ and $a_\lambda$ are bosonic creation and annihilation operators for polarization $\lambda$, and $k_0$ is the photonic wavevector (assumed to be on resonance with the dipole transition). In the rotating wave approximation, the dipole-field interaction Hamiltonian can be written as

$$H_{\text{int}} = -\sum_{\text{j}=1}^{N} \mathbf{p}_j \cdot \mathbf{E}(z_j) = -g\left(S_{k_0-p}^- a_-^\dagger + S_{k_0+p}^- a_+^\dagger + \text{H.c.}\right), \quad (3)$$

where $g = \wp\mathcal{E}/\sqrt{2}$ and $S_q^\pm = \sum_j \exp(\pm iqz_j)\sigma_j^{eg/ge}$ are collective raising and lowering operators for the giant coherent dipole with wavevector $q$ (H.c. stands for Hermitian conjugate). For the nanometer-scale dipole separations present in our superlattices, $k_0 \ll p$ and therefore $H_{\text{int}}$ manifests a chiral photonic spin-orbit coupling where modes with opposite circular polarization have opposite quasimomentum and group velocity. Accounting for room-temperature dephasing and tracing out the field degrees of freedom in the usual way (see Methods for details), one can derive a set of coupled differential equations for the mode intensities $I_{\text{LCP}}(t)$ and $I_{\text{RCP}}(t)$.

The solid lines in Fig. 3d show the theoretically calculated $\text{DCP}_0$ for the SMBA (red) and RMBA (blue) perovskites. The theoretical results show excellent agreement with the experimental data, capturing both the superlinear amplification of circular polarization at low pump fluence as well as saturation at high pump fluence. Here, the dephasing rate $\gamma_\phi$ and the initial mode imbalance $I_{\text{LCP}}(0)/I_{\text{RCP}}(0)$ are fit parameters (see Methods). The dephasing rates used in Fig. 3d ($1/\gamma_\phi =$ 1.7 ps for SMBA and 2.5 ps for RMBA) are roughly consistent with the experimentally estimated $T_2$ times measured at high pump fluence (Supplementary Note 8), whereas the initial mode imbalance is a free parameter but required by symmetry to yield a nonzero circular polarization during chiral SF[9,10,38]. Such an imbalance may be traced back to the forward-backward asymmetry imbued by the growth substrate and/or the top layer orientation of the fluorescent edge modes (Supplementary Note 4). As such, our theoretical results support our experimental observations that chiral SF results from the interplay between quantum coherence and structural chirality, rather than any intrinsic circular polarization of individual emitters.

**Magnetic field-tunable chiral SF**

In our chiral superlattices, a weak external magnetic field $B < 0.5$ T can be used to enhance and modulate the chiral SF response. As illustrated in Fig. 4a, magneto-optical measurements were performed on each perovskite sample with the magnetic field oriented perpendicular to the SF detection direction. When pumped just below the SF threshold, chiral SF in the SMBA and RMBA perovskites can be switched on or off via the magnetic field, transitioning from unpolarized spontaneous emission ($B = 0$ T) to circularly polarized SF ($B = 0.4$ T) (Fig. 4b, Extended Data Fig. 8). This behavior is visualized in contour maps of the emission intensity recorded under a periodically varying magnetic field (see Extended Data Fig. 9 for the opposite field direction). The ability to switch chiral SF on and off via the magnetic field underscores its high stability and external controllability.

The observed magnetic field dependence of chiral SF is summarized in Fig. 4c. The upper panel shows the LCP SF intensity for the SMBA superlattice under above-threshold excitation. The difference between the chiral SF intensity at magnetic field $B$ and at zero field, $I_{SF}(B)-I_{SF}(0)$, grows quadratically with the magnetic field up to ~0.5 T. A similar quadratic scaling is observed for the RMBA superlattice (Supplementary Fig. 26). The DCP also increases quadratically with the field strength, reaching up to 16.5% for the SMBA superlattice (lower panel, Fig. 4c; see SF spectra in Supplementary Fig. 27). By contrast, the unpolarized SF from the achiral PEA superlattice shows no such magnetic field-dependent behavior within the 0–0.5 T range (Fig. 4b, Supplementary Fig. 26). We speculate that the magnetic field may act as a selective amplifier for chiral SF by isolating the energy levels of the circularly polarized collective states. Such a mechanism could result in an increase in coherent population buildup by reducing transitions to other states, thereby reducing the effective SF threshold and enhancing the SF intensity and DCP. Alternatively, mixing of collective states by the perpendicular magnetic field may differentially modify the SF decay rates for the two circular polarizations, further amplifying the effect. Although a deeper theoretical investigation is required, the contrasting response between the chiral and achiral superlattices suggests a fruitful future platform for exploring collectively enhanced chiral light-matter interactions.

**Conclusion**

Our work demonstrates room temperature chiral SF from large-area, vertically aligned chiral quasi-2D perovskite superlattices. The effect originates from a spontaneous phase coherence among helically aligned dipoles and enables chiral SF emission with up to ~14% circular polarization. The lack of circularly polarized spontaneous emission in the chiral perovskites at room temperature highlights the essential role of cooperative light–matter interactions in amplifying the chiral response. Our theoretical calculations corroborate these experimental findings and establish the fundamental mechanism underlying chiral SF as a photonic chiral spin-orbit coupling between collective dipolar modes. We further show that a weak external magnetic field can significantly enhance both the chiral SF intensity and the degree of circular polarization, demonstrating the exceptional stability and tunability of these systems. These findings advance our fundamental understanding of chirality in quantum-optical phenomena and open new pathways for quantum spin-optical applications, including chiral light sources and scalable quantum information architectures.

## References from main text

## Main figure legends

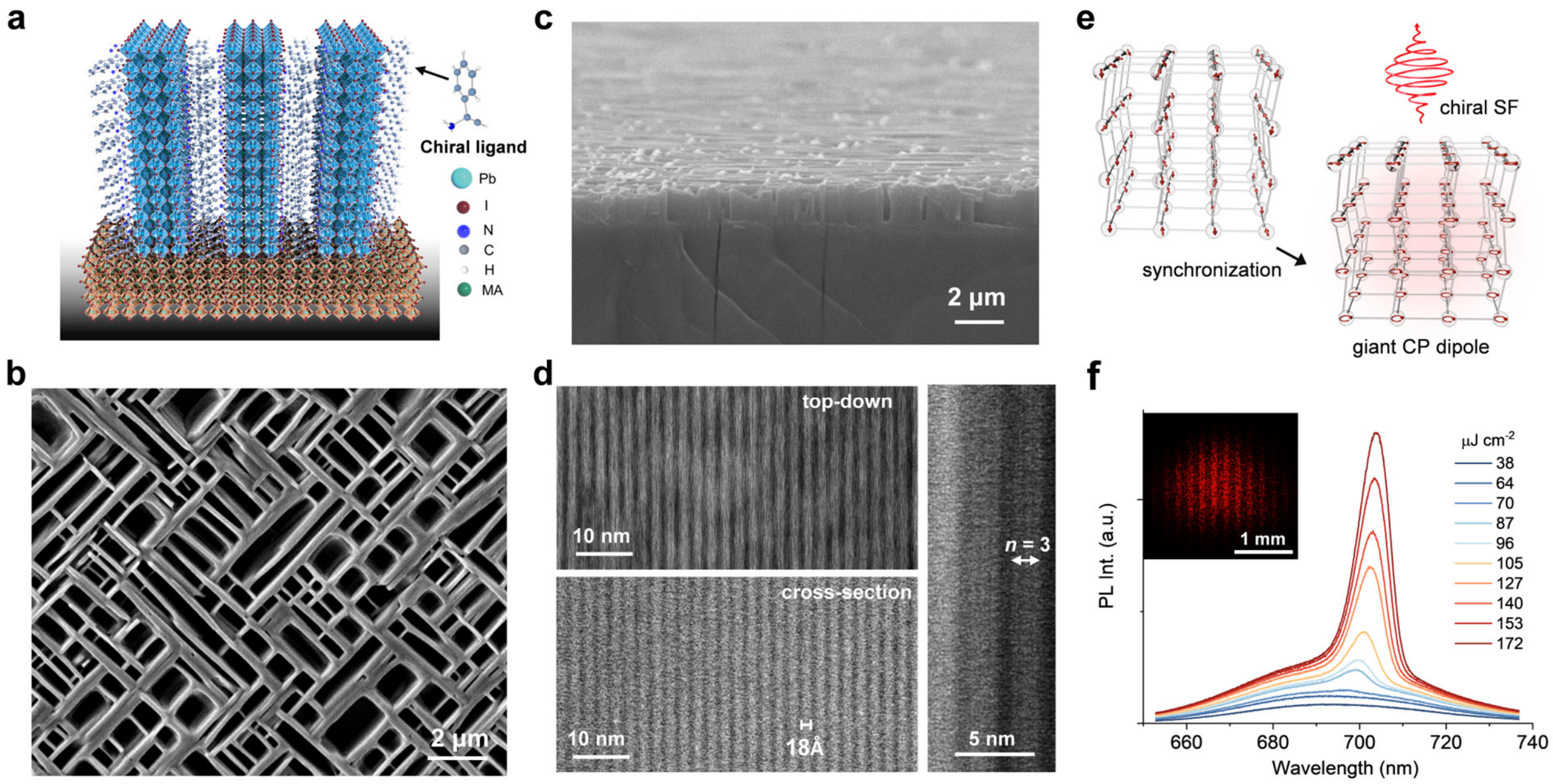

**Fig. 1 | Structure and spectra of chiral perovskite superlattices. a,** Schematic diagram of quasi-2D perovskite superlattices grown vertically on a $MAPbBr_3$ substrate. **b,** Top-view and **c** cross-sectional view scanning electron microscope images. **d,** Top-down and cross-section STEM images of the SMBA (left-handed) chiral perovskite superlattices. The number of octahedral layers $n$=3 in each quantum well is shown on the right of **d** with a thickness of 18Å. **e**, Schematic illustration of the spontaneous formation of a giant circularly polarized dipole from an initially incoherent dipolar ensemble, leading to chiral SF. Small arrows indicate individual dipole phasors. **f**, Power-dependent PL spectra of the SMBA perovskite superlattice under 550 nm linearly polarized pump excitation at room temperature. Inset, the real-space interferogram images collected by a camera using a Michelson interferometer above the SF threshold. Scale bars: 2 μm (**b**,**c**); 10 nm (**d**, left top and bottom); 5 nm (**d**, right), 1 mm (**f**, inset).

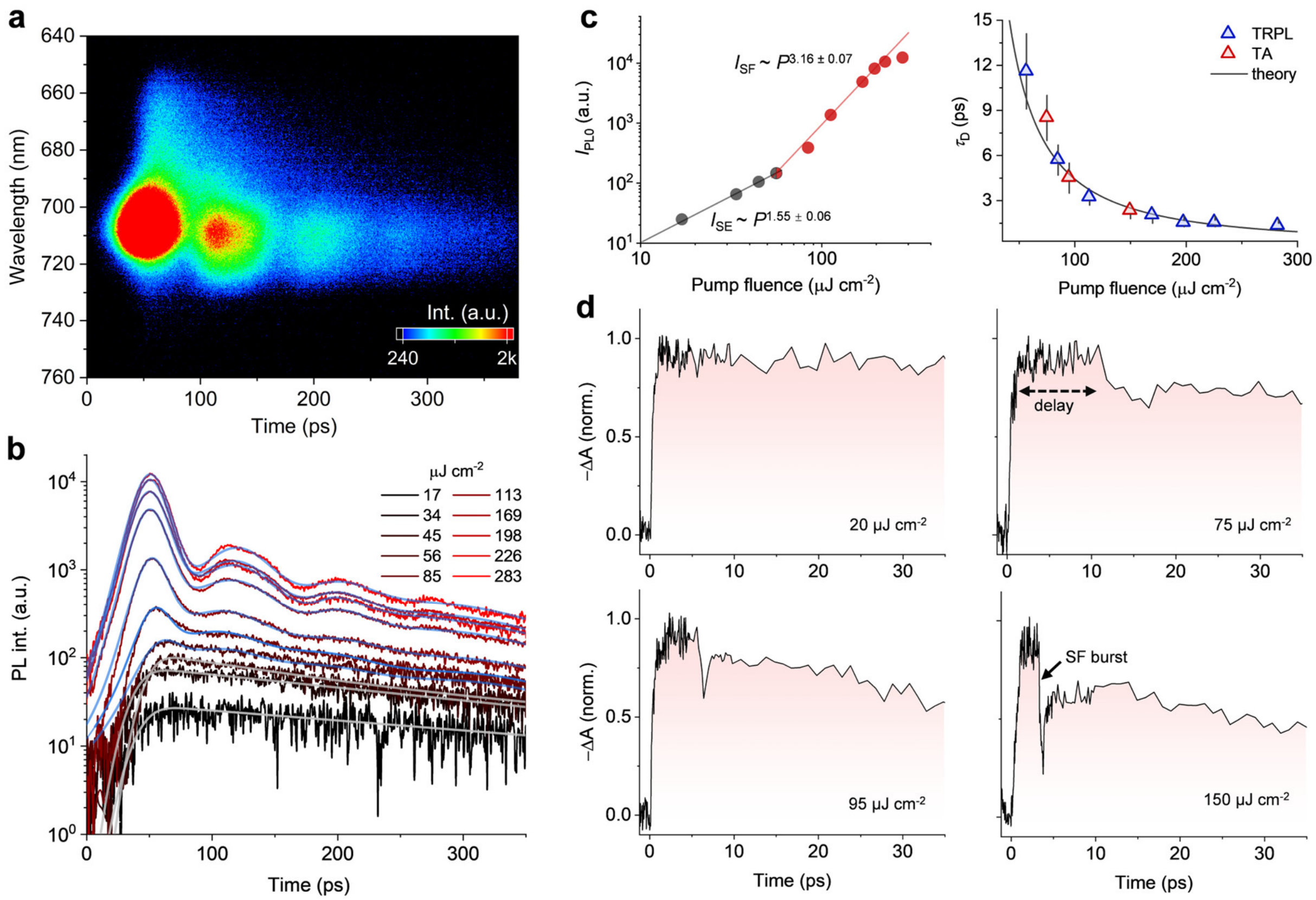

**Fig. 2 | Superfluorescence signatures and dynamics.** **a**, Streak camera image of the SMBA perovskite superlattices under 550 nm excitation, pumped above the SF threshold (198 μJ cm⁻²). The color bar denotes the emission intensity in arbitrary units. **b**, Pump fluence dependent time-resolved PL spectra. The light blue lines for SF dynamics show a fit to the sum of individual SF pulses in order to account for the ringing effect (Supplementary Note 1.2). The gray lines are single exponential decay fittings for spontaneous emission below the SF threshold. **c**, Left, the peak SF emission intensity as a function of pump fluence. Solid lines show individual power-law fits for spontaneous emission (SE) and SF. The errors in scaling parameters represent fitting uncertainties. Right, the delay time $\tau_D$ extracted from time-resolved PL and from transient absorption (TA) spectra, plotted as a function of pump fluence. The solid curve shows a fit to $\tau_D = (1/\Gamma)\ln(N)/(N+1)$, where $\Gamma = 1/1500$ ps$^{-1}$, $N = bP^{1.55}$, and $b$ is a fit parameter. **d**, Normalized transient absorption dynamics probed at the 705 nm peak SF position of the SMBA perovskite superlattices.

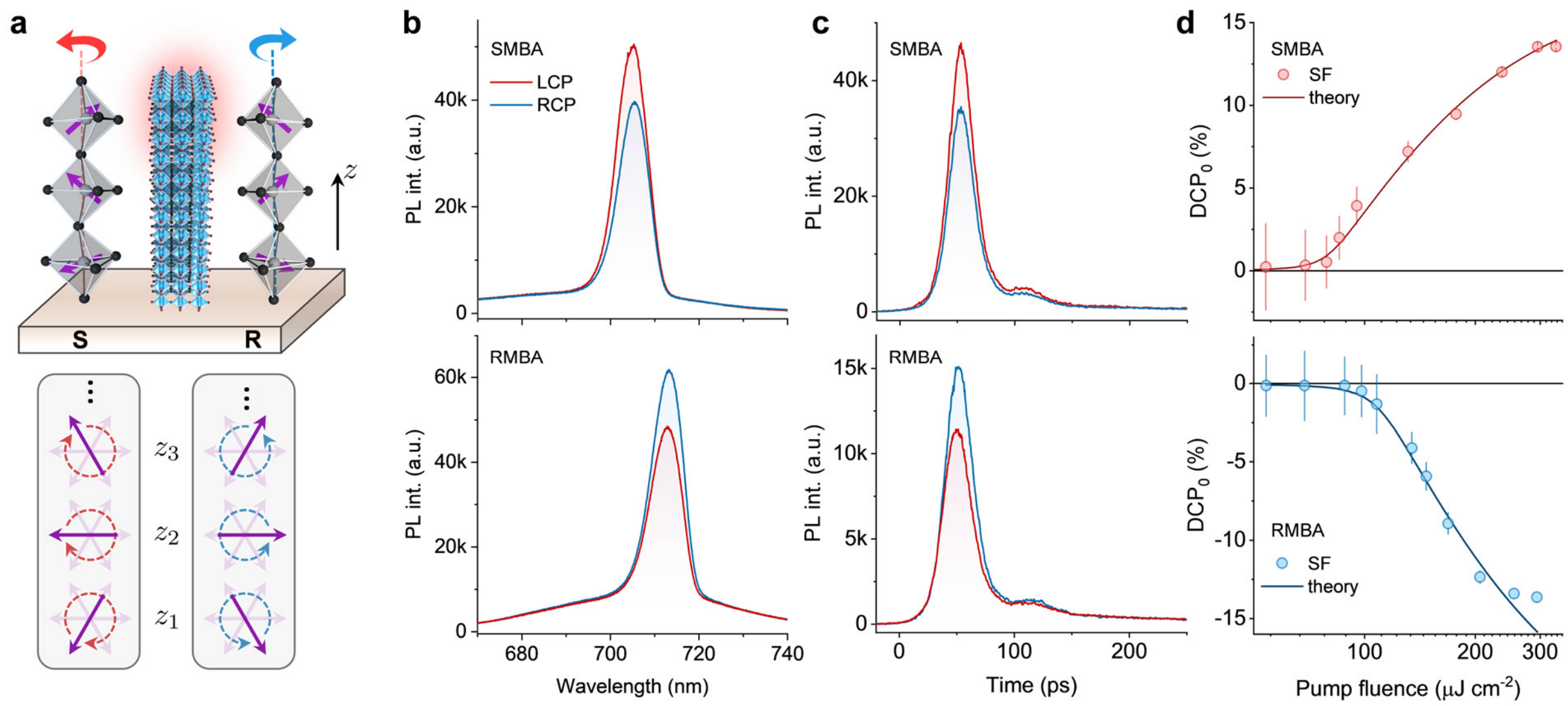

**Fig. 3 | Circularly polarized chiral superfluorescence. a**, Schematic of chiral SF via edge states in a chiral perovskite superlattice (top panel) with collectively rotated dipoles in the z-direction (lower panel). **b**, Time-integrated and **c,** time-resolved PL spectra showing left- and right-circularly polarized SF from chiral (S/R)MBA perovskite superlattices under linearly polarized excitation at 550 nm with a pump fluence of 300 μJ cm$^{-2}$. **d**, Degree of circular polarization at peak emission ($DCP_0$) as a function of pump fluence. The solid curves are theoretical calculations. Error bars represent the propagated uncertainty from the determination of TRPL peak intensities.

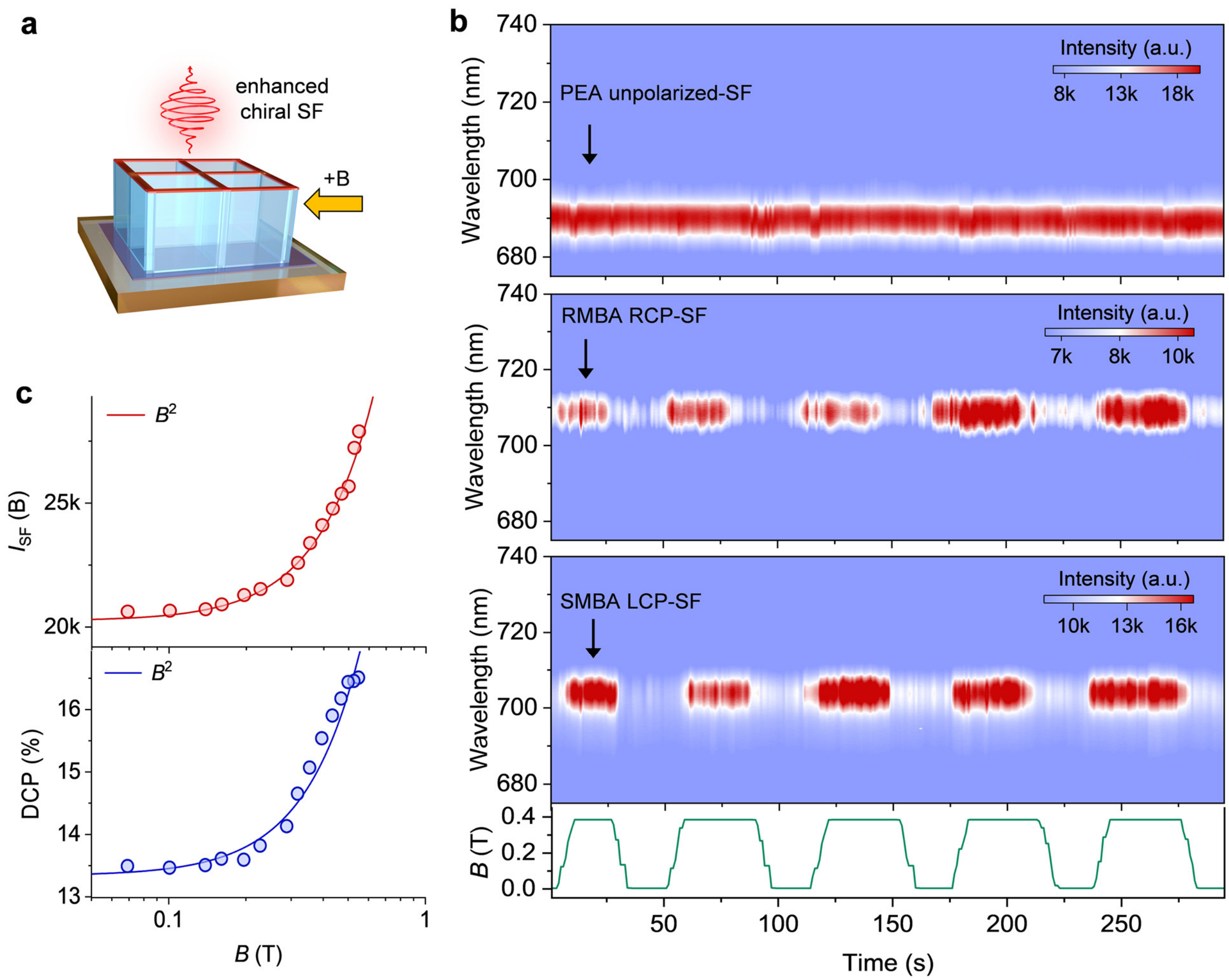

**Fig. 4 | Magnetic field-tunable chiral superfluorescence**. **a**, Sketch of the perovskite superlattices illustrating the magneto-spectroscopy scheme. The samples are driven by a 550 nm laser pulse and subject to a magnetic field oriented perpendicular to the SF detection direction. **b**, 2D color plot of PL spectra for achiral PEA (170 μJ cm⁻²), left-handed SMBA (68 μJ cm⁻²) and right-handed RMBA (92 μJ cm⁻²) perovskite superlattices as a function of periodic magnetic field strength. **c**, Top: Magnetic field dependence of the left-handed circularly polarized SF intensity for the SMBA perovskite superlattice (198 μJ cm⁻²). The solid curve is a fit to $I_{\mathrm{SF}}(B) = I_{\mathrm{SF}}(0) + b'B^2$. Bottom: Corresponding time-integrated degree of circular polarization (DCP). The solid curve is a fit to $\mathrm{DCP}(B) = \mathrm{DCP}(0) + c'B^2$, where $b'$ and $c'$ are fitting parameters.

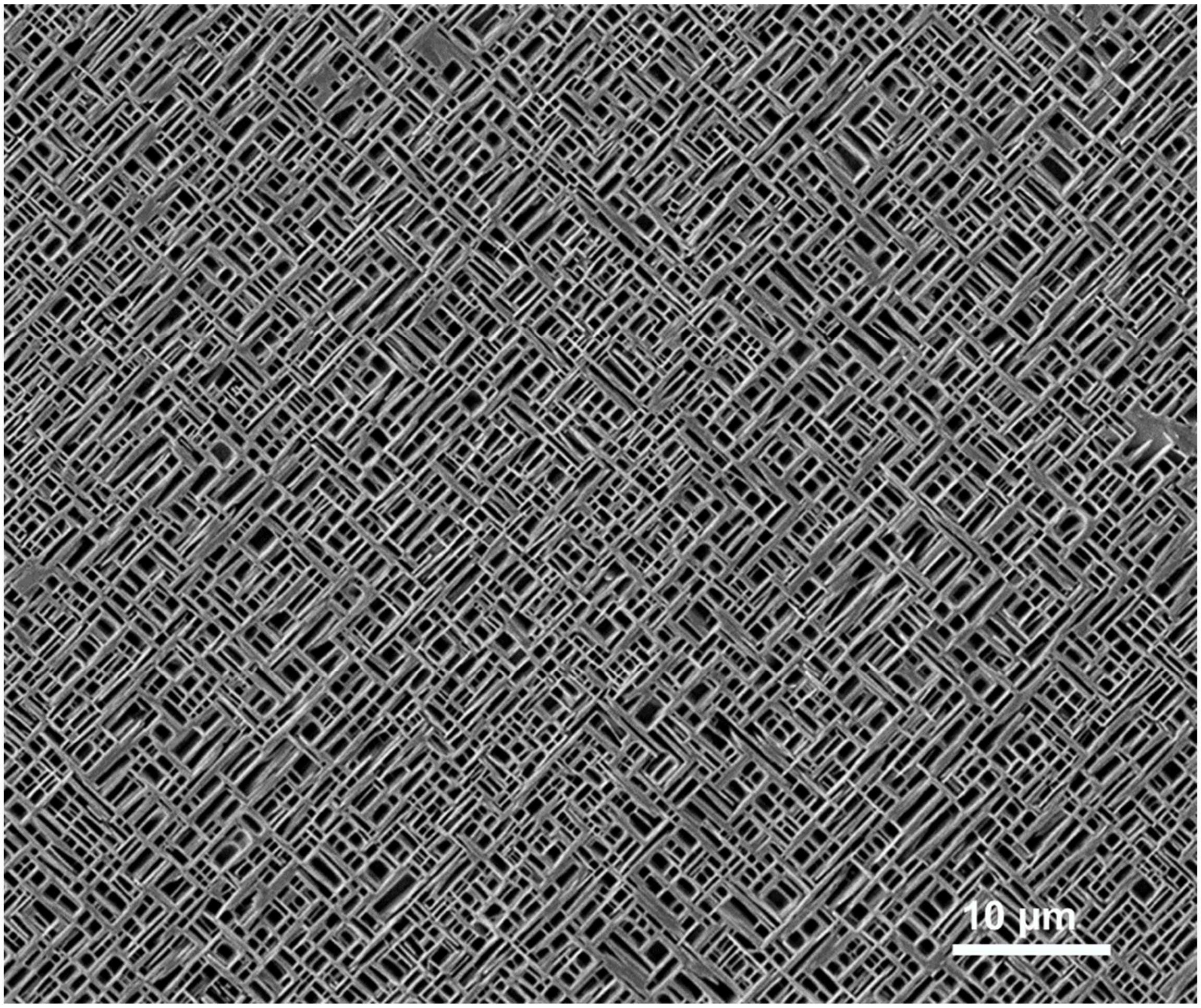

**Extended Data Fig. 1 | Large-area vertically aligned quasi-2D perovskite superlattices.** Top-view scanning electron microscope image of chiral perovskite (SMBA) quantum well superlattices grown vertically on the $MAPbBr_3$ single-crystal substrate.

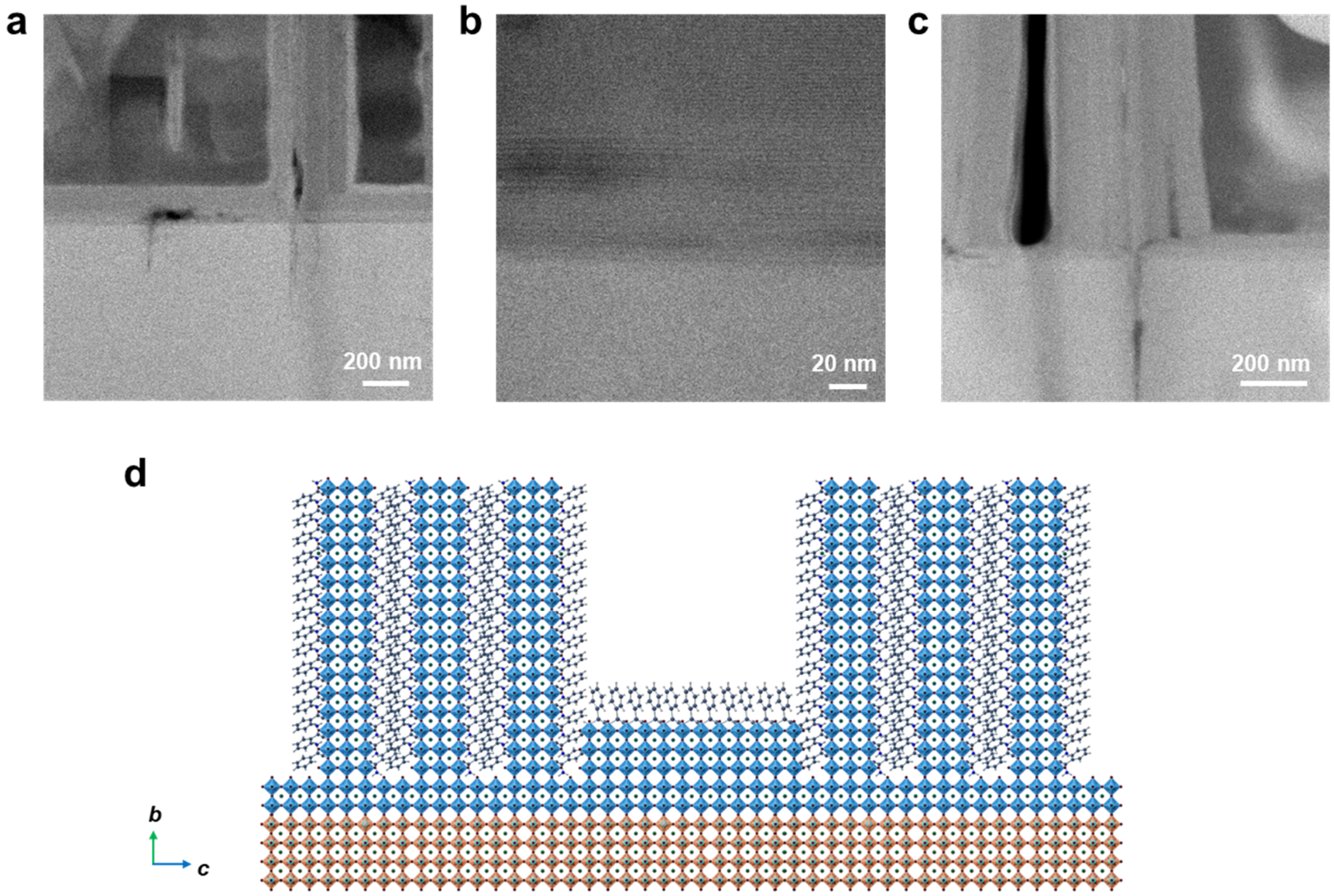

**Extended Data Fig. 2 | Cross-sectional view of quasi-2D perovskite superlattices. a-c**, Cross-sectional STEM images at interfaces of chiral perovskite superlattices on $MAPbBr_3$ single crystal substrate. **d**, Schematic cross-section view of perovskite superlattices on $MAPbBr_3$ single crystal substrate, where the horizontal perovskites (shown in the middle) correspond to the holes of the crisscross structured samples.

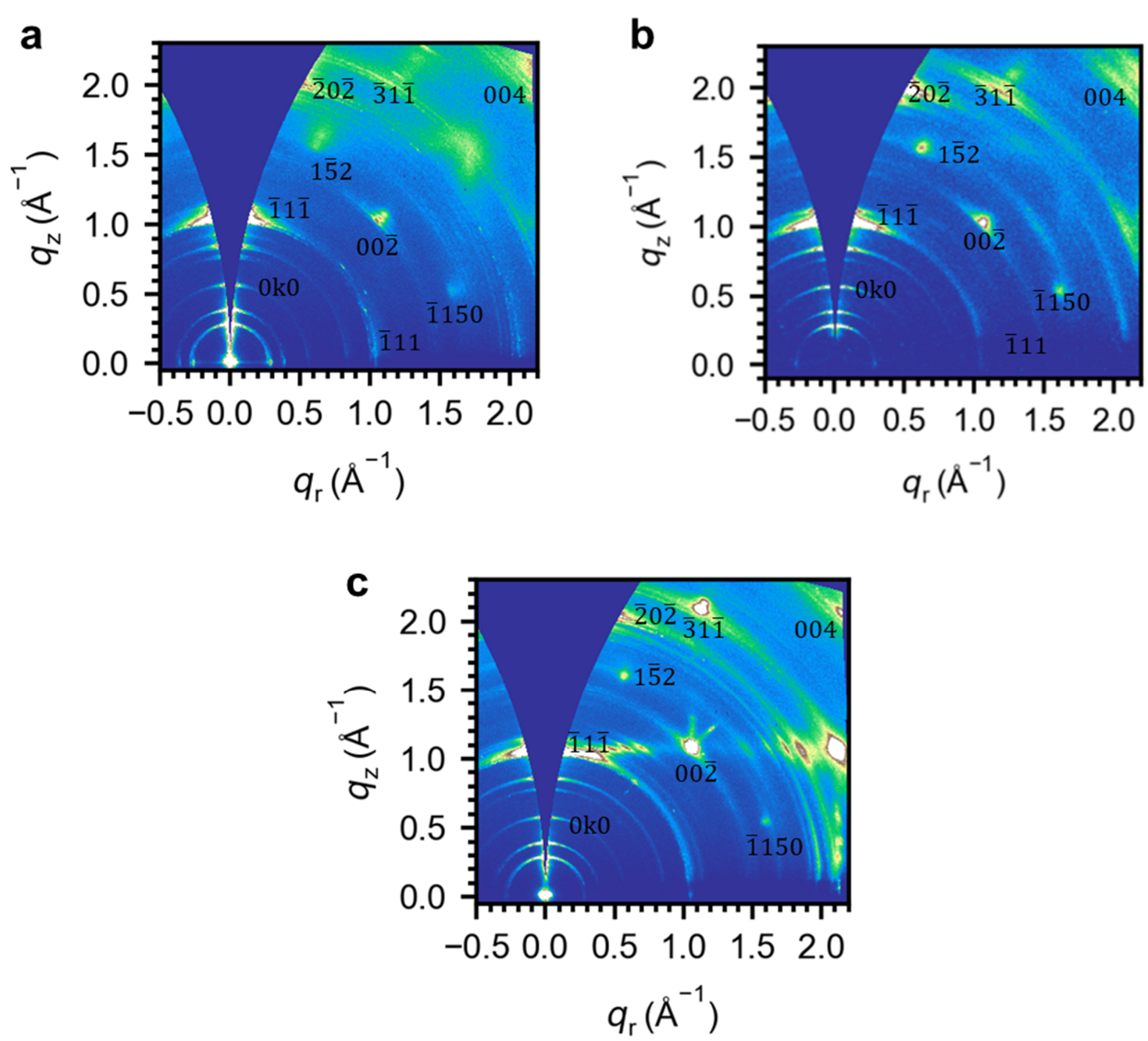

**Extended Data Fig. 3 | Perovskite superlattice structures.** Synchrotron-based grazing incidence wide angle scattering (GIWAXS) spectra of **a**, SMBA, **b**, RMBA, and **c,** PEA perovskite superlattices. The incident angle was set to 0.7°. All the superlattices show clear streak patterns, indicating vertical alignment. The isolated small-angle (< 1 Å$^{-1}$) streaks in the $q_{xy}$ and $q_z$ directions indicate the long-range ordered superlattice structure.

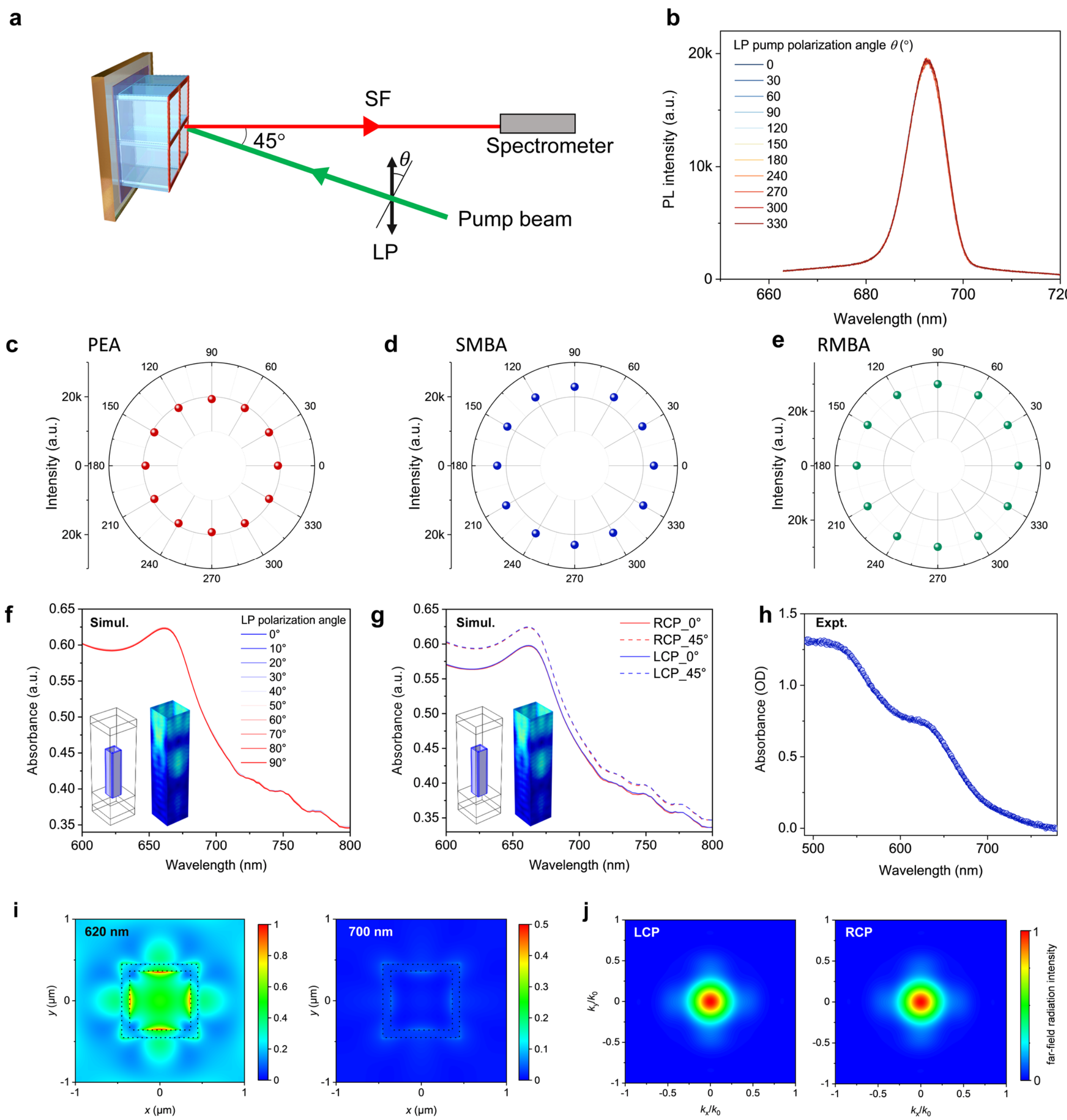

**Extended Data Fig. 4. | Photonic properties of perovskite superlattices. a**, Schematic of the experimental setup for SF measurement with a linearly polarized (LP) pump incident at 45° relative to the substrate normal. **b**, SF spectra of achiral PEA perovskite superlattices under illumination with different pump polarization angles. **c–e**, SF peak intensity as a function of pump polarization angle for PEA, SMBA, and RMBA superlattices. **f**, Simulated absorption spectra for the superlattice structure at a 45° incident angle for different linear polarization directions. **g**, Simulated absorption spectra for LCP/RCP light at normal and 45° incident angles. Insets in f and g show the simulation structure and the corresponding electric field distributions within the superlattices. **h**, Measured absorption spectrum of an SMBA superlattice transferred onto a quartz substrate. **i**, Simulated top-down view of near-field

electric field distributions in the superlattice, indicating the absence of waveguiding to the edge for emissions from bulk with wavelengths of 620 nm and 700 nm. **j**, Simulated far-field radiation intensity distributions for an RCP and an LCP dipole within the superlattice structure, which are indistinguishable. The simulations are consistent with experimental data and confirm that the observed circularly polarized chiral SF from edge states is not a result of photonic effects related to the morphologies of the superlattices. See Supplementary Note 3 for details of simulation methods.

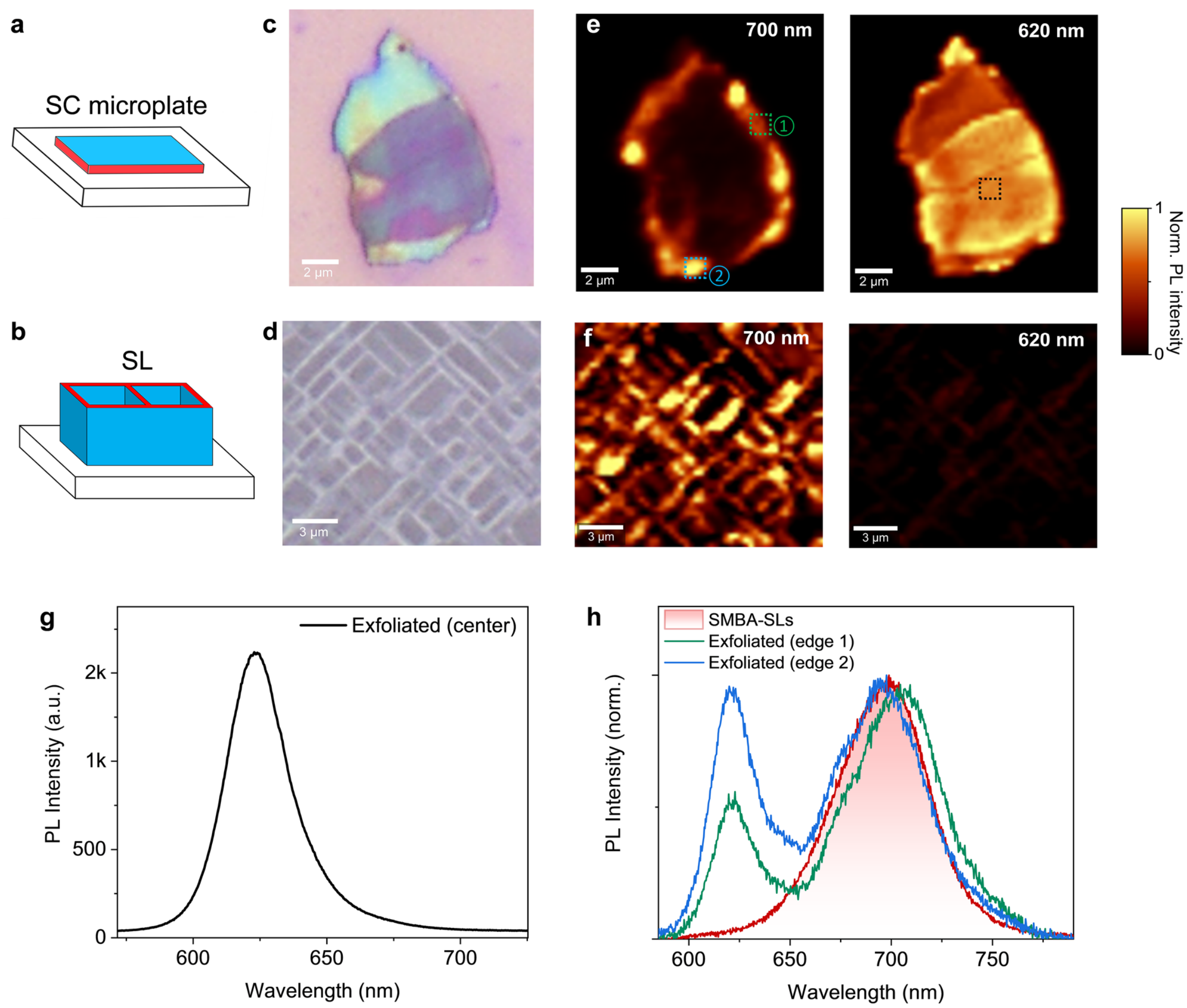

**Extended Data Fig. 5 | PL mapping of exfoliated single-crystal microplate and superlattice.** Schematic representation of **(a)** an exfoliated single-crystal (SC) microplate and **(b)** a superlattice (SL), with edge states highlighted in red. Top-view optical images of **(c)** an exfoliated microplate (control sample) from a $(PEA)_{1.2}(S\text{-}MBA)_{0.8}(MA)Pb_3I_{10}$ ($n$ = 3) single crystal and **(d)** an SMBA superlattice. PL mapping images at 620 nm (left) and 700 nm (right) for the **(e)** microplate and **(f)** superlattice, respectively. **(g)** PL spectra taken at the center (black dashed box in **e**, right) and **(h)** at the edge positions (green dashed box in **e**, left) of the microplate, compared with the PL spectra of the superlattice.

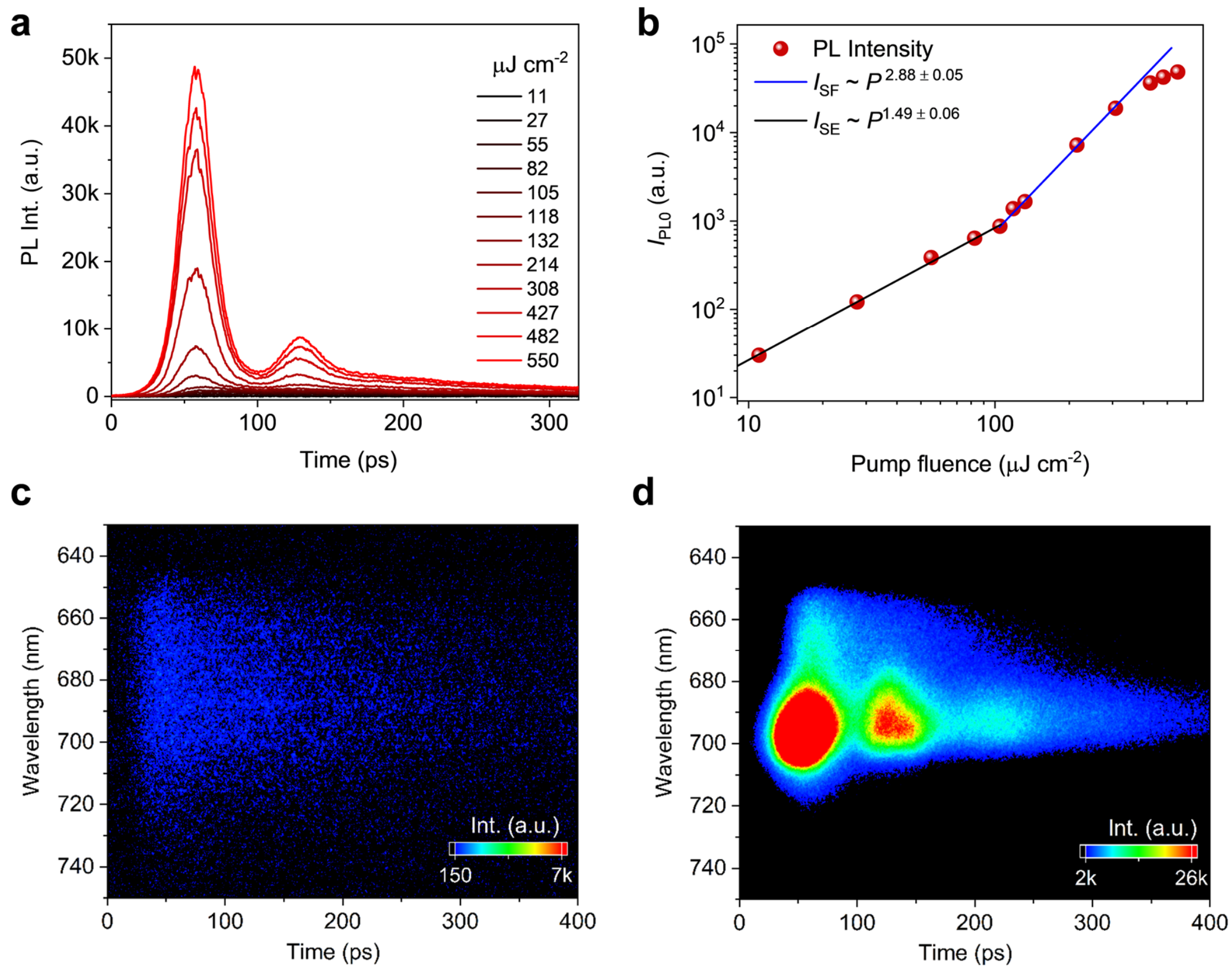

**Extended Data Fig. 6 | Burnham-Chiao ringing of PEA perovskite superlattices**. **a** Power-dependent time-resolved photoluminescence spectra. **b** Peak emission intensity ($I_{PL0}$) versus pump fluences. The solid lines are power-law fits in the regions of spontaneous emission (black) and SF (blue), respectively. Streak camera images below (55 μJ cm$^{-2}$) **c** and above (427 μJ cm$^{-2}$) **d** the SF threshold, respectively.

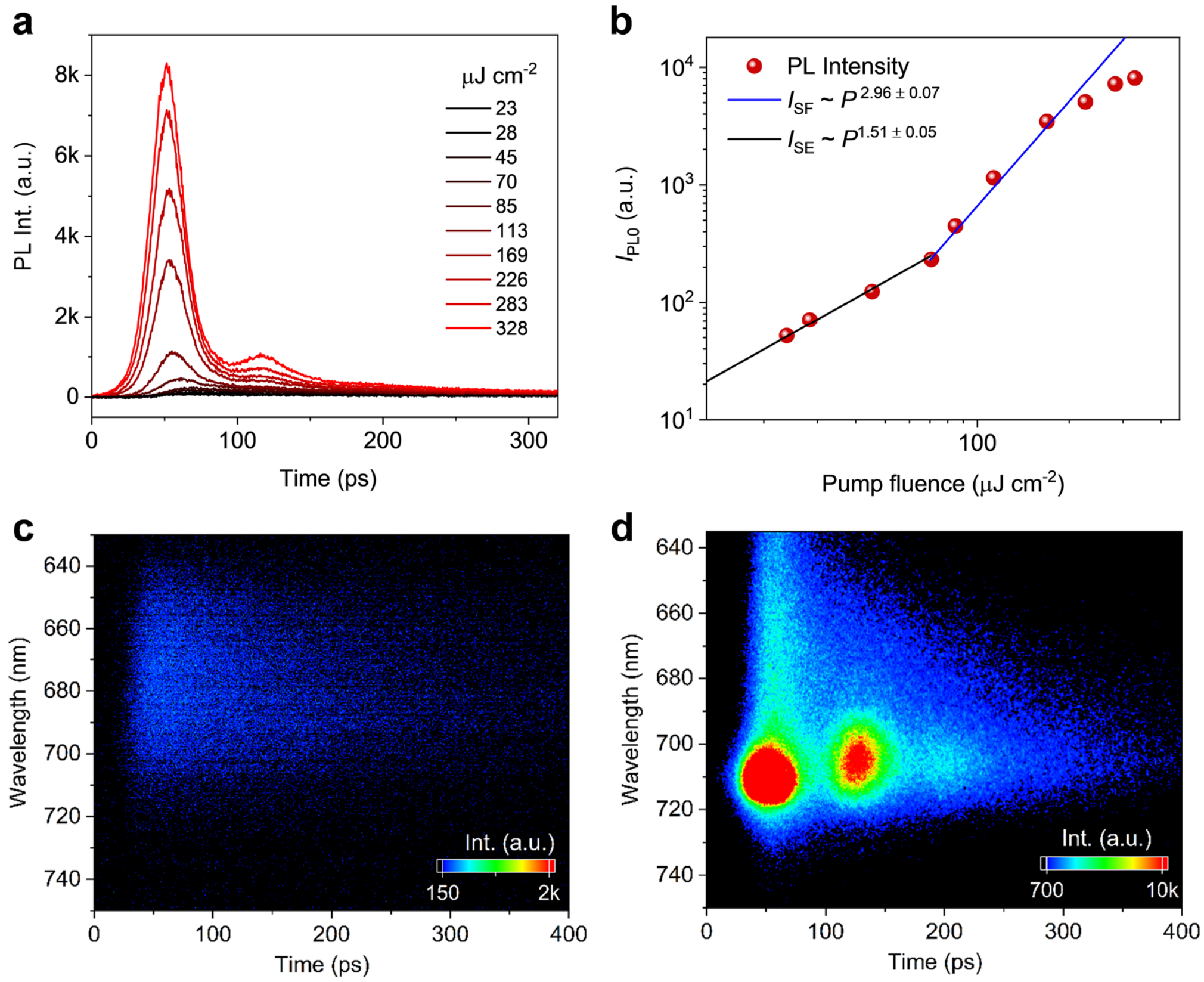

**Extended Data Fig. 7 | Burnham-Chiao ringing of RMBA chiral perovskite superlattices**. **a** Power-dependent time-resolved photoluminescence spectra. **b** Peak emission intensity ($I_{PL0}$) versus pump fluences. The solid lines are power-law fits in the regions of spontaneous emission (black) and SF (blue), respectively. Streak camera images below (45 μJ cm$^{-2}$) **c** and above (328 μJ cm$^{-2}$) **d** the SF threshold, respectively.

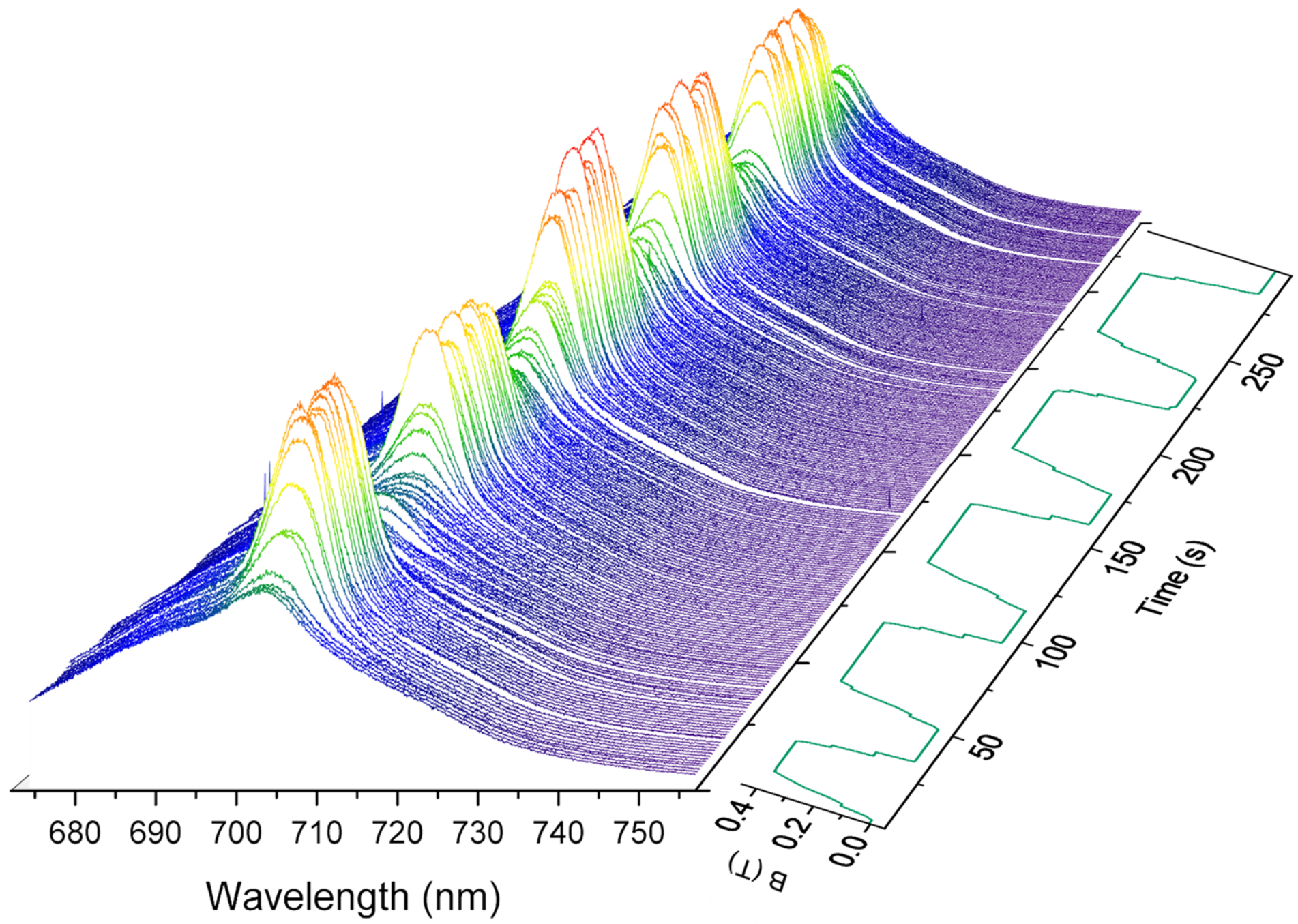

**Extended Data Fig. 8 | Magnetic field tunable photoluminescence spectra of SMBA chiral perovskites.** The sample is pumped just below the zero-field SF threshold (68 μJ cm$^{-2}$). Color bars indicate the intensity of the emission. Green, yellow, and red contours indicate narrow-band SF when the magnetic field (B) is turned on.

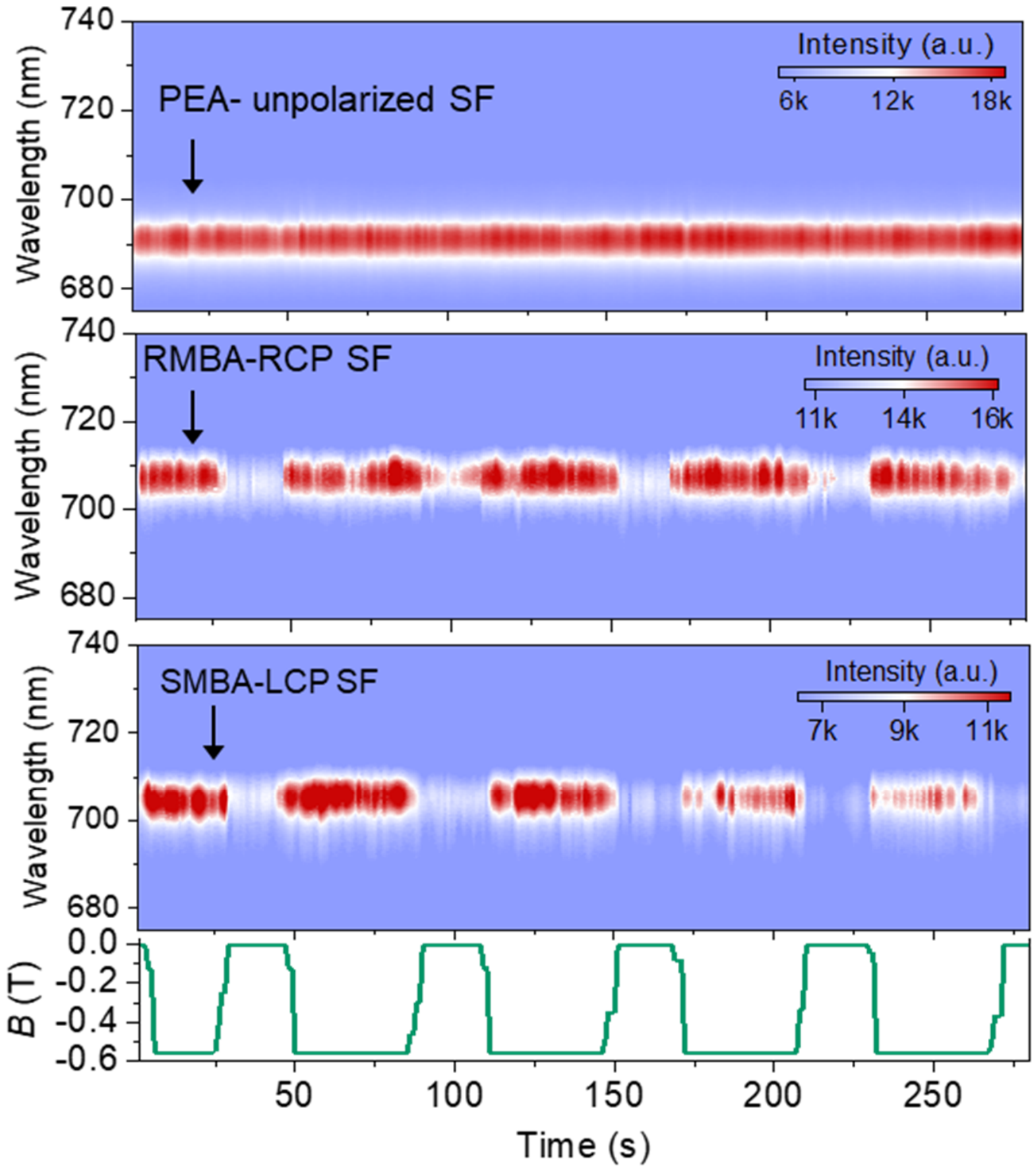

**Extended Data Fig. 9 | Magnetic field tunable chiral superfluorescence of perovskite superlattices.** The magnetic field direction is opposite to the one in Fig. 4b.